\documentclass[10pt,twocolumn]{article}

\usepackage[a4paper,margin=0.72in,columnsep=0.24in]{geometry}
\usepackage[T1]{fontenc}
\usepackage[utf8]{inputenc}
\usepackage{lmodern}
\usepackage{microtype}
\usepackage{booktabs}
\usepackage{tabularx}
\usepackage{array}
\usepackage{amsmath}
\usepackage{amssymb}
\usepackage{graphicx}
\usepackage{xcolor}
\usepackage{enumitem}
\usepackage{xspace}
\usepackage{url}
\usepackage{hyperref}
\usepackage{balance}
\usepackage{tikz}
\usetikzlibrary{arrows.meta,positioning,fit,backgrounds}

\definecolor{ink}{HTML}{14213D}
\definecolor{blue}{HTML}{2563EB}
\definecolor{cyan}{HTML}{0891B2}
\definecolor{green}{HTML}{15803D}
\definecolor{amber}{HTML}{B45309}
\definecolor{softblue}{HTML}{EAF1FF}
\definecolor{softcyan}{HTML}{E7F7FA}
\definecolor{softgreen}{HTML}{ECF8EF}
\definecolor{softamber}{HTML}{FFF4E5}
\definecolor{rulegray}{HTML}{CBD5E1}

\hypersetup{
  colorlinks=true,
  linkcolor=blue,
  citecolor=blue,
  urlcolor=cyan,
  pdfauthor={Yeongjin Jo},
  pdftitle={Secret MCP: Evidence-Bounded and Context-Isolated Design Specification Generation from Web Screenshots}
}

\setlist{nosep,leftmargin=*}
\newcommand{\system}{\textnormal{\textsc{Secret MCP}}\xspace}
\newcommand{\code}[1]{\texttt{\detokenize{#1}}}
\newcommand{\measured}{\textnormal{\textsc{Measured}}\xspace}
\newcommand{\observed}{\textnormal{\textsc{Observed}}\xspace}
\newcommand{\inferred}{\textnormal{\textsc{Inferred}}\xspace}
\newcommand{\unknown}{\textnormal{\textsc{Unknown}}\xspace}

\title{\vspace{-1.1em}\textbf{Secret MCP: Evidence-Bounded and Context-Isolated\\Design Specification Generation from Web Screenshots}}
\author{Yeongjin Jo\\
Independent Researcher\\
\href{mailto:appsky1888@gmail.com}{appsky1888@gmail.com}}
\date{August 24, 2026}

\begin{document}
\raggedbottom
\maketitle

\begin{abstract}
Screenshot-to-code systems optimize for a rendered implementation, but screenshots omit the document structure, interaction logic, responsive rules, and provenance needed to distinguish observation from guesswork. Multi-reference prompts introduce an additional risk: evidence or inferred decisions from one reference can contaminate another. We present \system, an open-source local system that produces one auditable, implementation-oriented design specification per public web reference. It separates retrieval, evidence preparation, model invocation, storage, and inspection. Long desktop captures are resized to at most 1,200 pixels wide and split into 1,600-pixel tiles with 80-pixel overlap. Evidence records retain prepared- and source-space coordinates and an eight-color measured palette. A 19-section contract requires page inventories, navigation geometry, section bounds, responsive matrices, components, accessibility requirements, acceptance criteria, and explicit labels for \measured, \observed, \inferred, and \unknown claims. References are processed sequentially through a sampler interface. The evaluated Model Context Protocol adapter sends one \code{sampling/createMessage} request per reference with \code{includeContext: none}; a fresh-process adapter is available when a stronger execution boundary is required.

We evaluate public commit \code{c130c9c} at two levels. A live retrieval and fixture-model integration run selected two references after excluding a third, prepared nine evidence images, issued two distinct sampling requests, and produced two documents with zero cross-reference identifier occurrences. A static audit of three bundled, externally generated design indexes found all 19 required sections in every document, one unique reference identifier per document, eight page specifications, 1,943 pixel-valued measurements, and 319 color literals. These tests establish orchestration invariants and syntactic contract compliance, not semantic or visual reconstruction accuracy. The sampler abstraction preserves the same boundaries across direct model APIs and future transports despite the 2026 deprecation of MCP sampling.
\end{abstract}


\section{Introduction}

Generating frontend code from an image is now a well-established multimodal task. Early systems mapped screenshots to domain-specific language tokens \cite{beltramelli2018pix2code}; later work learned simplified markup from screenshots \cite{lee2023pix2struct}, introduced large synthetic corpora \cite{laurencon2024websight,yun2024web2code}, and developed real-world benchmarks with visual, element, layout, and capability-level metrics \cite{si2025design2code,guo2025iwbench,lin2025webuibench}. These efforts expose a recurring difficulty: a screenshot contains pixels, not the source document object model, CSS, interaction logic, responsive rules, accessibility semantics, or design rationale. A plausible rendering can therefore conceal unsupported assumptions.

The problem becomes harder when a user requests several references at once. A single multimodal context can contain screenshots, metadata, and instructions for multiple works. The model may accidentally transfer a color, route, component, or inferred interaction from one work to another. Even when the final pages look reasonable, a user has little evidence for determining which claim came from which image. A monolithic response also makes retries expensive and complicates partial failure recovery.

\system addresses this workflow problem rather than proposing a new vision or language model. It creates a provenance-preserving intermediate artifact, named \code{DESIGN_INDEX}, between screenshot evidence and implementation. Each design reference is processed separately, its evidence and request contract are stored, and its document is inspectable without loading another work's document body. The approach is model-agnostic: the server specifies what must be measured and disclosed, while the connected model supplies the analysis.

This paper makes four contributions:

\begin{enumerate}
  \item a reference-scoped generation architecture that maintains a one-reference, one-request, one-document invariant and rejects combined fallback when the configured boundary is unavailable;
  \item an evidence preparation scheme for long screenshots that records overlapping tile coordinates, scale mappings, compressed image bytes, and quantized color measurements;
  \item a 19-section specification contract that converts visual observations into page-scoped geometry, components, responsive behavior, accessibility requirements, and falsifiable acceptance criteria while exposing uncertainty; and
  \item a reproducibility evaluation covering build health, live retrieval, exclusion behavior, request isolation, persisted evidence, and syntactic contract compliance of bundled outputs.
\end{enumerate}

The evaluation is deliberately narrow. We do not claim that section counts, coordinate literals, or identifier isolation prove visual fidelity. They verify system invariants and output structure. Controlled human evaluation and rendered-page similarity experiments remain future work.

\section{Related Work}

\subsection{Screenshot-to-code generation}

\textit{pix2code} demonstrated end-to-end generation of platform-specific interface tokens from a single screenshot \cite{beltramelli2018pix2code}. Pix2Struct subsequently used masked screenshot parsing into simplified HTML as a pretraining objective for visually situated language understanding \cite{lee2023pix2struct}. WebSight released two million synthetic HTML--screenshot pairs and fine-tuned a vision-language model for HTML generation \cite{laurencon2024websight}, while Web2Code combined instruction-tuning data with webpage understanding and code-generation evaluation \cite{yun2024web2code}.

Recent benchmarks broaden the evaluation target. Design2Code uses 484 real-world webpages and combines visual similarity with text, position, color, and block matching \cite{si2025design2code}. IW-Bench evaluates element completeness and relative layout over 1,200 image--code pairs \cite{guo2025iwbench}. WebUIBench separates perception, HTML programming, screenshot--HTML understanding, and end-to-end generation across more than 21,000 question--answer pairs \cite{lin2025webuibench}. Nguyen et al. isolate a pattern-completion bias: models can recognize an anomalous visual element yet still replace it with the repeated, pattern-consistent value \cite{nguyen2026pattern}.

These systems primarily target executable code or benchmark model capabilities. \system instead targets an explicit intermediate specification and the orchestration boundary around its generation. This distinction matters when evidence is incomplete or conflicts with learned layout regularities: a code generator must choose a value, while a specification can state whether a value is measured, directly visible, inferred for implementation, or unknown. The artifact can then guide a separate implementation model, developer, or evaluation loop. We do not compare model quality against these benchmarks in the present study.

\subsection{Protocol-mediated model invocation}

The Model Context Protocol (MCP) introduced client capability negotiation and server-initiated \code{sampling/createMessage} requests. The request includes messages, a required maximum token count, model preferences, and an optional context-inclusion value \cite{mcp2025sampling}. \system uses the capability as an adapter for delegating one analysis back to the host model.

MCP specification version 2026-07-28 deprecated sampling while retaining wire-level behavior during a transition window; SEP-2577 identifies direct provider integration as an alternative for servers that require model access \cite{vangent2026sep2577}. This paper therefore treats MCP sampling as the evaluated transport, not as an essential scientific assumption. The system's generator accepts a sampler callback whose input is one reference, one contract, and one evidence list. A direct API adapter can preserve the same request and storage invariants.

\section{Problem Formulation}

Let a retrieval procedure produce ordered references
\(
R = \langle r_1, \ldots, r_n \rangle
\).
Each reference \(r_i\) contains metadata \(m_i\) and a set of source screenshots \(S_i\). Evidence preparation produces \(E_i = P(S_i)\), where each evidence item includes image bytes, a coordinate transform, and palette measurements. Contract construction produces \(C_i = C(m_i,E_i)\). A sampler \(G\) returns markdown \(D_i = G(C_i,E_i)\).

The system aims to preserve the following invariants:

\begin{align}
\textbf{I1: } & \quad \operatorname{refs}(C_i \cup E_i) = \{r_i\}, \\
\textbf{I2: } & \quad \operatorname{history}(G_i) \cap \{D_j: j<i\} = \varnothing, \\
\textbf{I3: } & \quad |\{G_i\}| = |\{D_i\}| \quad \text{for successful items}, \\
\textbf{I4: } & \quad D_i \text{ is stored before preparation of } r_{i+1}, \\
\textbf{I5: } & \quad A_i=(m_i,C_i,E_i,D_i,L_i) \text{ is inspectable},
\end{align}

where \(L_i\) is the per-item event trace. I1 prevents mixed request evidence. I2 defines the desired context boundary. With the MCP adapter, \code{includeContext: none} requests this property, but the protocol permits the client to constrain or modify request parameters; a fresh process and workspace are therefore needed when execution-level non-reuse is required. I3 and I4 make partial failures local and observable. I5 supports audit rather than relying only on the final prose.

\begin{figure*}[t]
\centering
\resizebox{0.98\textwidth}{!}{%
\begin{tikzpicture}[
  font=\sffamily\small,
  node distance=7mm and 8mm,
  box/.style={draw=ink,rounded corners=2pt,align=center,minimum height=10mm,inner xsep=7pt,fill=white},
  input/.style={box,fill=softblue},
  evidence/.style={box,fill=softcyan},
  model/.style={box,fill=softamber},
  artifact/.style={box,fill=softgreen},
  arrow/.style={-{Latex[length=2.2mm]},thick,draw=ink}
]
\node[input] (user) {User query\\limit and filters};
\node[input,right=of user] (search) {GDWEB search\\persistent exclusions};
\node[evidence,right=of search] (prepare) {Per-reference evidence\\resize, tile, palette};
\node[evidence,right=of prepare] (contract) {19-section contract\\coordinates and labels};
\node[model,right=of contract] (sample) {Sampler adapter\\one isolated request};
\node[artifact,right=of sample] (store) {Atomic per-work save\\contract, evidence, document};
\node[artifact,right=of store] (viewer) {Manifest and viewer\\one selected work};
\draw[arrow] (user) -- (search);
\draw[arrow] (search) -- node[above,font=\scriptsize]{\(r_i\)} (prepare);
\draw[arrow] (prepare) -- node[above,font=\scriptsize]{\(E_i\)} (contract);
\draw[arrow] (contract) -- node[above,font=\scriptsize]{\(C_i,E_i\)} (sample);
\draw[arrow] (sample) -- node[above,font=\scriptsize]{\(D_i\)} (store);
\draw[arrow] (store) -- (viewer);
\draw[arrow,rounded corners=5pt] (store.south) -- ++(0,-7mm) -| node[pos=0.25,below,font=\scriptsize]{next work only after save} (prepare.south);
\node[draw=blue,dashed,rounded corners=3pt,fit=(prepare)(contract)(sample)(store),inner sep=4mm,label={[blue]below:reference-scoped sequential boundary}] {};
\end{tikzpicture}}
\caption{\system execution pipeline. Search is shared, but evidence preparation, contract construction, model invocation, and storage operate on one reference at a time. The loop advances only after the current artifact is persisted.}
\label{fig:architecture}
\end{figure*}
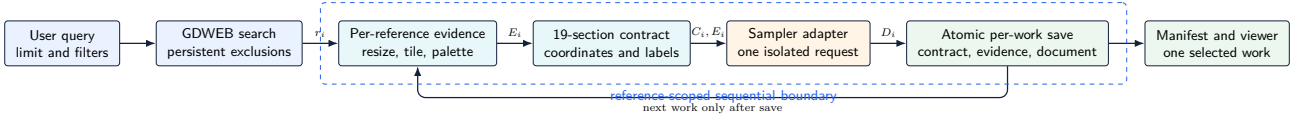

\section{System Design}

\subsection{Retrieval and exclusion}

The primary generation tool accepts a query, a limit from one to ten, year and award filters, language, output directory, and a per-work token budget. The retrieval module searches GDWEB metadata and reads more candidates when exclusions would otherwise reduce the requested result count. Exclusions are persistent but prospective: excluding a reference affects future searches and does not delete prior evidence or documents. The active exclusion set is copied into each run manifest.

This design keeps user curation outside the language-model prompt. Filtering occurs before evidence loading and before any model request. A lightweight list-only tool exists, but generation does not expose a list to the outer host and ask it to compose a combined analysis. Instead, the server retains the selected items and constructs the sequence internally.

\subsection{Long-image evidence preparation}

GDWEB source captures may be extremely tall. Passing an original multi-megabyte image directly can exceed transport limits and can make small layout details difficult for a vision model to resolve. \system normalizes each desktop screenshot as follows:

\begin{enumerate}
  \item auto-rotate and flatten transparency on white;
  \item downscale without enlargement to width \(W_p \leq 1200\) pixels;
  \item JPEG-encode at quality 78;
  \item split the prepared canvas into tiles of height at most 1,600 pixels with 80-pixel vertical overlap; and
  \item record both prepared-space and source-space rectangles for each tile.
\end{enumerate}

Mobile images are kept as separate evidence and are not merged with desktop tiles. For a prepared crop with top \(y_p\), height \(h_p\), and vertical scale \(s_y=H_p/H_s\), source mapping is

\begin{equation}
\begin{aligned}
  y_s &= \operatorname{round}(y_p/s_y), \\
  h_s &= \min\!\left(H_s-y_s,\operatorname{round}(h_p/s_y)\right).
\end{aligned}
\end{equation}

The 80-pixel overlap preserves context at tile boundaries. The contract explicitly declares overlap as duplicated evidence and requires sections to be deduplicated by prepared-canvas coordinates.

\begin{figure}[t]
\centering
\resizebox{\columnwidth}{!}{%
\begin{tikzpicture}[font=\sffamily\scriptsize,x=0.75mm,y=0.75mm]
  \fill[softblue] (0,0) rectangle (38,70);
  \draw[ink,thick] (0,0) rectangle (38,70);
  \node[rotate=90] at (-4,35) {source screenshot \(H_s\)};
  \draw[blue,thick] (1,51) rectangle (37,69);
  \draw[cyan,thick] (1,34) rectangle (37,53);
  \draw[green,thick] (1,17) rectangle (37,36);
  \draw[amber,thick] (1,1) rectangle (37,19);
  \node[fill=white,inner sep=1pt] at (19,60) {tile 1};
  \node[fill=white,inner sep=1pt] at (19,43.5) {tile 2};
  \node[fill=white,inner sep=1pt] at (19,26.5) {tile 3};
  \node[fill=white,inner sep=1pt] at (19,10) {tile 4};
  \draw[<->,draw=ink] (41,51) -- (41,53);
  \node[anchor=west] at (43,52) {80 px overlap};
  \draw[-{Latex},thick,draw=ink] (54,35) -- (72,35);
  \node[align=left,anchor=west] at (76,50) {Per tile:};
  \node[align=left,anchor=west] at (76,42) {prepared \((x,y,w,h)\)};
  \node[align=left,anchor=west] at (76,34) {source \((x,y,w,h)\)};
  \node[align=left,anchor=west] at (76,26) {scale \((s_x,s_y)\)};
  \node[align=left,anchor=west] at (76,18) {8 measured colors};
\end{tikzpicture}
}
\caption{Schematic evidence tiling. Overlap supplies continuity; offsets map measurements to prepared and source canvases.}
\label{fig:tiling}
\end{figure}
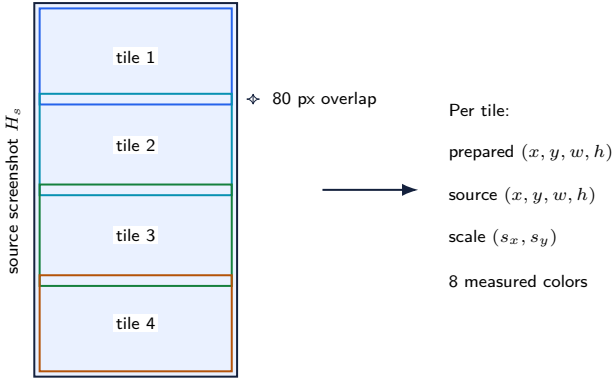

\subsection{Measured palette extraction}

Each tile is resized to fit within a \(180\times180\) sample, flattened, and converted to raw RGB. Channels are quantized to multiples of 17, yielding \(16^3=4096\) possible bins. The eight most frequent bins are stored with HEX, RGB, HSL, and pixel coverage. If \(N\) sampled pixels are present and bin \(b\) contains \(n_b\) pixels, coverage is \(n_b/N\), rounded to four decimals.

These values are objective measurements of normalized screenshot pixels, not recovered CSS tokens. Photographic content, antialiasing, scaling, and JPEG compression all affect frequency. The contract therefore requires a sampled value to be labeled \measured, while a proposed semantic CSS variable is \inferred\ unless independently confirmed.

\subsection{The DESIGN\_INDEX contract}

The contract uses schema identifier \code{secret-mcp/design-index/v2} and defines four evidence labels. \measured claims are derived from supplied coordinates or palette samples. \observed claims are directly visible or present in supplied metadata, but are not numerically measured. \inferred claims record a concrete implementation decision required to reproduce the evidence. \unknown claims are not recoverable from static evidence and must not be presented as fact.

The required 19 sections are grouped in Table~\ref{tab:contract}. The contract is intentionally detailed because a design index is meant to be consumed without reopening the source gallery. Every visible route gets its own canvas model, ordered section table, components, state rules, responsive transitions, assets, accessibility requirements, and acceptance checks. A long scroll remains one page unless distinct route screens are visibly present.

\begin{table}[t]
\centering
\caption{Required DESIGN\_INDEX sections, grouped for readability.}
\label{tab:contract}
\small
\begin{tabularx}{\columnwidth}{@{}p{0.18\columnwidth}X@{}}
\toprule
Group & Required sections \\
\midrule
Scope & 1 Goal and scope; 2 evidence and coordinates; 3 route inventory \\
Shell & 4 application shell; 5 navigation and header \\
Pages & 6 page specifications; 7 section deep dives; 8 component abstraction \\
Visual & 9 tokens and colors; 10 typography; 11 assets and icons; 12 responsive matrix \\
Behavior & 13 interaction and motion; 14 accessibility; 15 data and content \\
Delivery & 16 frontend architecture; 17 task graph; 18 page acceptance; 19 uncertainties \\
\bottomrule
\end{tabularx}
\end{table}

\subsection{Reference-scoped generation}

The generator loop is sequential by construction. For each reference it loads source evidence, prepares images, saves the request contract and evidence, invokes the sampler, normalizes the returned markdown, writes the document, records model and timestamps, and only then advances. An empty response, non-text content, timeout, or token-limit stop reason marks only that item as failed.

The MCP adapter sends a system prompt, the contract, alternating metadata and image content blocks, temperature 0.2, model preferences, and a per-work maximum output budget. Version 0.6.0 accepts 131,072 to 262,144 tokens per work and treats a token-limit stop as failure rather than saving a known-truncated document. This range is a requested upper bound; the client may support a smaller practical output size. The high default reflects the worst-case multi-page contract and should not be interpreted as typical consumption.

At the protocol level, each call explicitly uses \code{includeContext: none}. This is a request to the client, not a cryptographic guarantee. For hosts without an acceptable sampling boundary, the documented direct client starts a new model process in a new temporary workspace per call and copies only that call's text and images. It does not reuse a conversation, response file, process, or working directory. Alternatively, a direct provider adapter can implement the same callback.

\subsection{Run manifest and viewer}

Each run directory contains \code{run.json}, one contract and one document per work, evidence images, and an event log. The manifest stores paths and status rather than concatenating document bodies. A local web viewer displays run progress and, for one selected work at a time, the specification, evidence and measurements, request contract, and generation log. Markdown is sanitized before browser rendering. Evidence and documents are read-only through the viewer; exclusion management is the only state-changing curation action.

\section{Evaluation}

\subsection{Research questions}

We evaluate three questions:

\begin{description}
  \item[RQ1] Can the public repository be installed, audited, built, and linted from a fresh clone?
  \item[RQ2] Does a live integration run preserve exclusion, one-request-per-reference, no-other-identifier, and per-item artifact invariants?
  \item[RQ3] Do the three bundled full design indexes syntactically satisfy the 19-section contract and remain reference-scoped?
\end{description}

RQ2 uses a deterministic fixture sampling handler. It evaluates orchestration and artifact structure, not the design-analysis capability of an external model. RQ3 is a static document audit and does not establish whether numerical values match the underlying screenshots.

\subsection{Environment and procedure}

We cloned the public \code{main} branch at commit \code{c130c9c80876ddf9594df9517174a163f48d523f}. The environment was Darwin 25.4.0 on ARM64, Node.js 24.11.1, and npm 11.6.2. The package requires Node.js 20.19 or later. We ran \code{npm ci}, \code{npm audit --audit-level=low}, \code{npm run build}, \code{npm run lint}, and \code{npm run smoke:gdweb-isolation}.

The smoke test first verifies rejection of a 32,000-token budget. It retrieves one live finance-related candidate, adds that reference to the output directory's exclusion store, then requests two eligible works from 2025 or 2026. The mock sampling client requires exactly one reference identifier, at least one image, \code{includeContext: none}, the 131,072-token default, and mandatory contract markers. It records request texts and returns a deterministic, reference-specific markdown fixture. After generation it verifies request cardinality, distinct identifiers, absence of the excluded identifier, absence of any other selected identifier in each request, per-document identity, artifact paths, evidence coordinates and palettes, and manifest-recorded exclusions.

For RQ3, \code{paper/evaluate_artifacts.mjs} scans the three generated documents committed under \code{docs/generated/food-godot-20260803}. It counts exact required section names, unique \code{gdweb-*} identifiers, page headings, evidence labels, pixel literals, and six-digit color literals. These are transparent lexical checks.

\subsection{Results}

All RQ1 checks passed. Dependency installation completed, the audit reported zero vulnerabilities, TypeScript compilation succeeded, and ESLint returned no errors or warnings.

Table~\ref{tab:smoke} summarizes RQ2. The live search selected \code{gdweb-26522} and \code{gdweb-24516} after excluding \code{gdweb-26905}. The two selected items produced two sampler requests and two output files. The first request contained four desktop tiles and one mobile image; the second contained three desktop tiles and one mobile image. The manifest recorded all nine images, source and crop coordinate fields, representative palettes, contracts, models, documents, and completion timestamps. No selected request text contained the other selected reference ID, and the excluded ID was never sampled. The run completed in approximately 2.85 seconds; this time is not a model-latency benchmark because the sampling handler is a local fixture.

\begin{table}[t]
\centering
\caption{Live isolation smoke test at commit \code{c130c9c}.}
\label{tab:smoke}
\small
\begin{tabularx}{\columnwidth}{@{}Xr@{}}
\toprule
Invariant or artifact & Result \\
\midrule
Requested eligible references & 2 \\
Active excluded references & 1 \\
Sampling requests & 2 \\
Generated documents & 2 \\
Failed items & 0 \\
Desktop tiles & 7 \\
Mobile evidence images & 2 \\
Total evidence images & 9 \\
Cross-reference ID occurrences in requests & 0 \\
Excluded reference sampled & No \\
Request context setting & \code{none} \\
Per-work budget & 131,072 tokens \\
\bottomrule
\end{tabularx}
\end{table}

Table~\ref{tab:audit} reports RQ3. Every document contained all 19 required section names and exactly one unique GDWEB identifier matching its filename. Across the three documents, the audit found eight page specifications, 3,579 lines, 48,154 whitespace-delimited words, 1,943 \code{px}-suffixed numeric values, and 319 six-digit color literals. The counts show that the outputs are detailed and contract-shaped; they do not prove that any measured-looking value is correct. The evidence label distribution also shows that generated documents contain more inferred statements than measured or observed statements, reinforcing the need to preserve labels rather than collapse them into a single certainty class.

\begin{table*}[t]
\centering
\caption{Static audit of bundled externally generated DESIGN\_INDEX documents. Label counts and literal counts are lexical. ``Sections'' is coverage of the 19 required titles.}
\label{tab:audit}
\small
\begin{tabular}{@{}lrrrrrrrrrr@{}}
\toprule
Reference & Pages & Lines & Words & Sections & Meas. & Obs. & Infer. & Unk. & px values & Colors \\
\midrule
gdweb-26387 & 6 & 1,020 & 14,643 & 19/19 & 93 & 59 & 432 & 56 & 556 & 77 \\
gdweb-26788 & 1 & 869 & 12,037 & 19/19 & 87 & 59 & 339 & 68 & 527 & 63 \\
gdweb-26853 & 1 & 1,690 & 21,474 & 19/19 & 156 & 60 & 809 & 67 & 860 & 179 \\
\midrule
Total & 8 & 3,579 & 48,154 & 57/57 & 336 & 178 & 1,580 & 191 & 1,943 & 319 \\
\bottomrule
\end{tabular}
\end{table*}

\subsection{What the evaluation establishes}

The results support the following limited claims. The repository is buildable in the stated environment. The evaluated orchestration emits one request and one artifact per successful reference, honors a prospective exclusion, records evidence metadata, and detects cross-reference identifiers in the tested fixtures. Bundled documents match the contract's required headings and identifier scope.

The results do not support claims of state-of-the-art screenshot understanding, pixel-perfect reconstruction, improved developer productivity, or semantic absence of contamination. A model could transfer a style without copying a reference identifier, and a document could contain an incorrect numeric value while passing the lexical audit. These questions require a labeled benchmark, rendered implementations, automated visual comparison, and human review.

\section{Discussion}

\subsection{Why an intermediate specification?}

Direct code generation is attractive because the result is immediately executable. An intermediate specification adds cost, but it separates three concerns that are otherwise entangled: what the screenshot supports, what an implementation needs, and what the model guessed. It also allows framework choice to remain downstream. The same page specification can guide React, static HTML, a game-engine web export, or manual development.

The artifact is especially useful for multi-page or long-scroll evidence. A model can reference stable page and section identifiers, a canonical coordinate system, and an explicit responsive matrix. A later implementation agent can be evaluated against acceptance criteria rather than a vague instruction to make the page ``look similar.''

\subsection{Isolation strength}

\system provides several layers of separation: pre-sampling exclusion, one-reference request construction, sequential generation, no prior document in subsequent messages, per-work paths, and a viewer that selects one document body. These layers reduce accidental mixing and make obvious failures observable.

However, \code{includeContext: none} alone is not a security boundary. A client controls model selection and may alter or ignore requested context behavior. A persistent model process could also retain hidden state outside visible messages. Execution-level isolation therefore requires a trusted adapter that launches a fresh process and workspace or a direct API call whose request is constructed solely from \((C_i,E_i)\). Even then, model weights contain general knowledge; the claim is absence of run-local cross-reference context, not absence of prior training information.

\subsection{Protocol evolution}

Sampling's 2026 deprecation is a practical limitation for the current MCP adapter. The core system need not be discarded. The \code{GdwebDesignIndexSampler} type is already a dependency-injected callback, and all request-specific material is assembled before invocation. Migration options include: (1) a direct provider client inside the server, (2) a direct external sampling client using standard input/output, or (3) a future multi-round-trip MCP pattern that returns an explicit model-input request to the client. Any migration should retain I1-I5, log the model and adapter version, and fail rather than silently combining works.

Direct provider integration changes the trust model. It gives the server control over request construction and context but also requires credentials, provider-specific retention review, streaming and retry logic, and explicit user consent. The current local architecture avoids embedding a provider credential in the server and lets the host choose the model. There is no universally superior choice; the paper's contribution is the invariant-preserving boundary around the adapter.

\subsection{Copyright and responsible use}

Screenshots, branding, copy, photography, and layout may be protected by copyright or trademark. \system reads images and metadata made public by a design-gallery site, but public availability does not imply permission to reproduce or redistribute a site. The contract instructs users to adapt copyrighted copy and brand assets unless authorized. The viewer is local, and run artifacts are not uploaded automatically.

The system should be used for analysis, interoperability, accessibility planning, or authorized reconstruction, not deceptive cloning. For publishable datasets or redistributed evidence, users must establish a license or other lawful basis, minimize retained images, and document provenance. The bundled code is MIT-licensed, but that license does not transfer rights in third-party screenshots.

\section{Threats to Validity and Limitations}

\textbf{Construct validity.} Cross-reference identifier counts are a deliberately narrow proxy for contamination: a model may transfer a visual pattern without copying an identifier. Likewise, required headings and numeric literals can be present even when content is wrong. The present measures establish observable orchestration and lexical properties, not semantic correctness. A stronger validator should check evidence links, coordinate consistency, page-inventory coverage, and executable acceptance criteria.

\textbf{Internal validity.} The integration test uses a deterministic fixture sampler. It validates request construction and artifact persistence without measuring model quality, cost, or latency. The live retrieval result also depends on the GDWEB state observed during the run. Repetition should record content hashes, fixed model and adapter versions, seeds where available, and complete timing and cost data.

\textbf{External validity.} The live boundary test contains two selected works and one excluded work, and the bundled audit contains three documents from one domain scenario. Results may not generalize to other galleries, languages, page lengths, route counts, or mobile coverage. A larger, rights-cleared frozen corpus should stratify these factors.

\textbf{No rendered reconstruction benchmark.} The strongest downstream test would implement each specification without reopening the source evidence, render the result at canonical viewports, and compare it with the reference using Design2Code- and IW-Bench-style visual, element, and layout metrics \cite{si2025design2code,guo2025iwbench}. A blinded developer study could compare direct screenshot-to-code prompting with specification-mediated implementation.

\textbf{Operational limitations.} Retrieval depends on GDWEB markup and image endpoints, so changes, rate limits, or removals can break live experiments. MCP sampling is also deprecated as of specification version 2026-07-28. Immediate priorities are a direct model-provider adapter, adapter-conformance tests, evidence content hashes, compression and tile-overlap ablations, perceptual color checks, and an evaluation of whether explicit \unknown labels reduce hallucinated behavior.

\section{Reproducibility and Availability}

The source code, English and Korean documentation, verification procedure, viewer, bundled design indexes, and audit script are available in the public repository \cite{secretmcp2026}. The evaluated revision is commit \code{c130c9c80876ddf9594df9517174a163f48d523f}. Core reproduction commands are:

\begin{verbatim}
npm ci
npm audit --audit-level=low
npm run build
npm run lint
npm run smoke:gdweb-isolation
node paper/evaluate_artifacts.mjs
\end{verbatim}

The live smoke test depends on current GDWEB availability. Its response is deterministic fixture text, so no external model credential is required. The arXiv package includes \code{main.tex}, \code{references.bib}, and generated \code{main.bbl}.

\section{Conclusion}

\system reframes multi-reference screenshot analysis as an evidence and isolation problem. It preserves recoverable coordinates, separates measurement from inference, requires a page-complete contract, and stores one request and artifact set per work. The evaluation confirms these orchestration and structural properties while leaving semantic and visual accuracy open. Its reference-scoped sampler interface also separates the enduring invariants from the deprecated MCP sampling transport, providing a practical baseline for auditable, model-agnostic design specification generation.

\begingroup
\scriptsize
\balance
\bibliographystyle{abbrv}
\bibliography{references}
\endgroup

\end{document}